\documentclass[aps,prl,preprint,amsmath,amssymb,showpacs]{revtex4-1}
\usepackage{graphicx}
\usepackage{dcolumn}
\usepackage{bm}
\usepackage{multirow}
\begin{document}

\title{High-$T_c$ $d^{0}$ ferromagnetism in a doped Mott insulator: the case of hydrogenated epitaxial graphene on SiC(0001)}

\author{Pengcheng Chen}
\affiliation{Department of Physics and State Key Laboratory of Low-Dimensional Quantum Physics, Tsinghua University, Beijing 100084, People's Republic of China }
\affiliation{Collaborative Innovation Center of Quantum Matter, Tsinghua University, Beijing 100084, People's Republic of China }
\author{Yuanchang Li}
\email{Electronic mail: liyc@nanoctr.cn}
\affiliation{National Center for Nanoscience and Technology, Beijing 100190, People¡¯s Republic of China}
\author{Wenhui Duan}
\email{Electronic mail: dwh@phys.tsinghua.edu.cn}
\affiliation{Department of Physics and State Key Laboratory of Low-Dimensional Quantum Physics, Tsinghua University, Beijing 100084, People's Republic of China }
\affiliation{Collaborative Innovation Center of Quantum Matter, Tsinghua University, Beijing 100084, People's Republic of China }
\affiliation{Institute for Advanced Study, Tsinghua University, Beijing 100084, People's Republic of China}
\date{\today}

\begin{abstract}
We show that the $d^{0}$ ferromagnetism with high Curie temperature ($T_c$) can be achieved in the electron doped hydrogenated epitaxial graphene on some certain SiC substrates through first-principles calculations. The pristine systems are found to be a Mott insulator independent of SiC polytypes (2$H$, 4$H$ or 6$H$) which, however, play a significant role in the modulation of magnetic interaction. Carrier doping enhances the ferromagnetic coupling due to the double exchange mechanism and thus realizes the phase transition from antiferromagnetism to ferromagnetism. A $T_c$ of around 400 K is predicted on the 2$H$-SiC. We employ a non-degenerate Hubbard model to demonstrate how the SiC affects the interfacial magnetism in intra-atomic Coulomb repulsion and intersite hopping interactions.
\end{abstract}

\maketitle
$d^{0}$ ferromagnet particularly with high Curie temperature ($T_c$) is a long-sought goal for its nontrivial magnetic mechanism and long spin relaxation time which would promise a better performance in future spintronic devices\cite{Coey2005660,spmag,han2014graphene}. In low-dimensional systems, the electron kinetic energy is decreased with the reduction of atom coordination number at the surface or interface having localized defect states, resulting in an enhanced Coulomb interactions/bandwidth ratio. This usually gives rise to a large density of states around the Fermi energy and thus the emergence of spin-splitting according to the Stoner criteria\cite{AStoner,BStoner}. Following such a paradigm, $d^{0}$ ferromagnets have been predicted based on 2$p$-electrons of carbon, oxygen and nitrogen at the interface, surface, edge or defects\cite{Bouzerar,Han,dev2008defect,slipukhina2011ferromagnetic,wu2010magnetism,martinez2010ferromagnetism,peng2009origin,pan2007room}. Basically, current semiconducting industry is established on the Si element. Thus the integration of $d^{0}$ ferromagnets with Si-based systems is also highly desired.

Recently, room temperature $d^{0}$ ferromagnetism was reported in hydrogenated epitaxial graphene on SiC(0001)\cite{XieAPL,GiesbersPRL}, yet the physical origin is still ambiguous, especially about the interpretations of local moments. Some researchers attributed it to the formation of unpaired electrons in graphene\cite{XieAPL} while others ascribed it to the interfacial Si dangling bond (DB) states\cite{GiesbersPRL}. This issue is also crucial from a practical point of view: if the former is true, an additional protection process has to be implemented to retain the unpaired electrons in the presence of air; otherwise, the graphene layer itself would offer an inherent protection. On the other hand, it is a bit surprising that little attention is paid to the SiC substrate used although its polarization effect was reported to be stacking sensitive, which naturally affects the local electrostatic environments at the interface, e.g., the on-site Coulomb repulsion\cite{arXivSiCdoping,Ristein}. In fact, different SiC polytypes (namely, 4$H$-SiC\cite{XieAPL} and 6$H$-SiC\cite{GiesbersPRL}) were employed in the experiments for the epitaxial growth of graphene.

In this work, we report the engineering of $d^{0}$ ferromagnetism in the hydrogenated epitaxial graphene on different polytypes of SiC through first-principles calculations, where the moment is associated with the interfacial Si-DB state. Importantly, we find a new route to tune the magnetic interactions between the local moments, namely, the stacking sequence of SiC bilayer. The pristine system, regardless of SiC stacking, is a Mott insulator and possesses an antiferromagnetic ground state but the magnetic coupling strength depends upon the stacking heavily. When the carrier density increases, the ferromagnetic coupling is enhanced because of the double exchange mechanism and an estimated $T_c$ about 400 K is achieved at an electron density of 4$\times$10$^{13}$/cm$^{2}$ on 2$H$-SiC. We employ the non-degenerate Hubbard model to demonstrate how the SiC stacking affects the magnetic properties of the system. Our work not only unravels the origin of ferromagnetism in hydrogenated epitaxial graphene but also reveals a feasible system that is integratable with the current semiconducting industry for the applications in spintronics.

First-principles calculations are performed using the projector augmented wave\cite{PAW} methods within the local spin density approximation (LSDA)\cite{CA} as implemented in the Vienna \emph{ab initio} simulation package (VASP)\cite{vasp}. The slab model is used with a $\sqrt{3}\times\sqrt{3}$\emph{R}30$^\circ$ supercell for all 2$H$-, 4$H$- and 6$H$-SiC (0001) surfaces, which can accommodate a 2$\times$2 graphene cell. The SiC substrate is modeled by justified four (2$H$, 4$H$) or six (6$H$) SiC bilayers with the H-terminated bottom surface. The SiC bilayers in the lower half are fixed at their respective bulk positions. All the other atomic positions are fully optimized using the conjugate gradient algorithm without any symmetry constraint until the residual forces become smaller than 0.01 eV/\AA. A Monkhorst-Pack $k$-point mesh of 12$\times$12$\times$1 was used for all Brillouin-zone integrations.

\begin{figure}[tbh]
  \centering
  \includegraphics[width=0.75\textwidth]{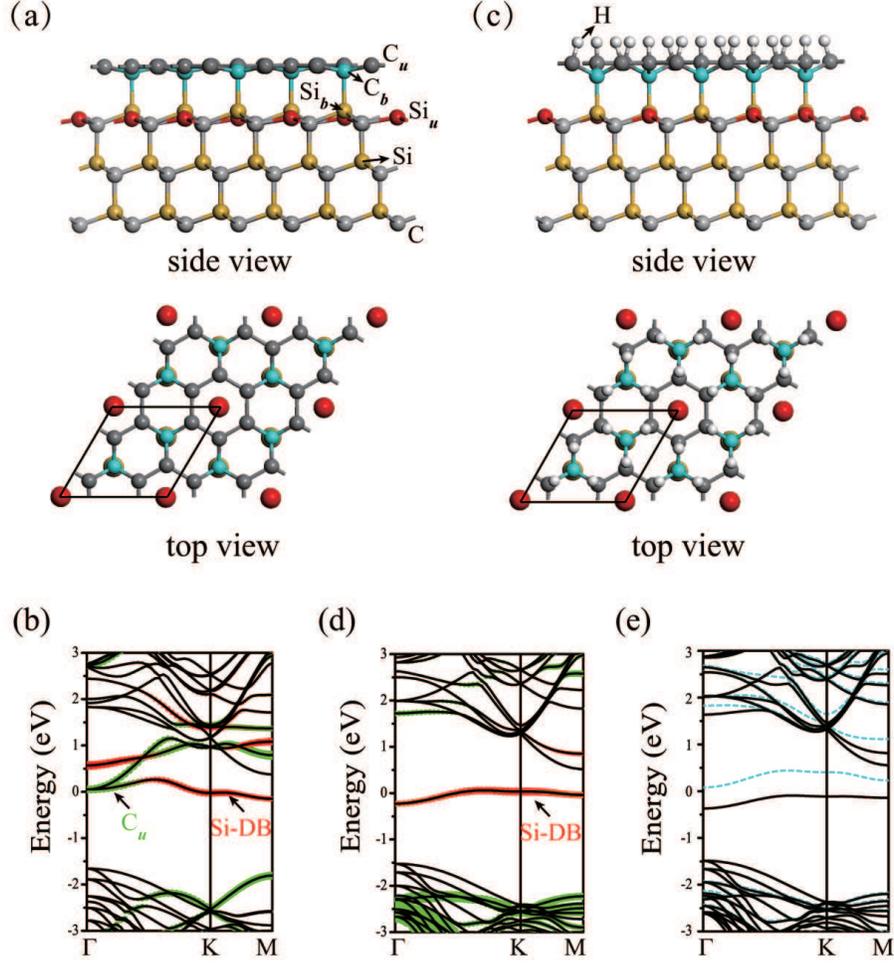}\\
  \caption{(Color online) (a) Side and top views of the first graphene layer on 6$H$-SiC(0001), where the subscripts of ``b'' and ``u'' respectively refer to the tetra- and triple-coordinated atoms, and (b) the corresponding band structure.  (c) Geometries after hydrogenation and the corresponding band structures without (d) and with (e) spin-polarization. Black rhombuses denote the surface $\sqrt{3}\times\sqrt{3}$ \emph{R}30$^\circ$ supercell in (a) and (c). The bands highlighted in red and green represents the states mainly contributed from Si$_u$ (Si-DB)and C$_u$ atoms, respectively. In (e), black solid and blue dashed lines correspond to the majority and minority spins, respectively. The Fermi level is set to zero.}\label{model}
\end{figure}

\emph{Local moments.}--For epitaxial graphene grown on SiC(0001), the first layer is so called ``buffer layer'', where a part of carbon atoms chemically bond to the Si atoms underneath. This results in the atomic corrugation\cite{Varchon,Mattausch,usprl,lijpcc}, which is typically illustrated in Fig. 1(a) by taking 6$H$-SiC as an example. It can be seen that two thirds of surface Si atoms bond to the carbon in the buffer layer while the remaining one third have the unsaturated DBs. The bonding between carbon and surface Si atoms violates the linearly dispersive band structure of graphene, leading to the metallic states contributed by Si$_u$ and C$_u$ around the Fermi level, as shown in Fig. 1(b). It is the hybridization between Si-DB and C$_u$ states that maintains the spin-degenerate states at the interface.

We then employ the hydrogenation approach to passivate the unsaturated C$_u$ states and decouple Si-DB and C$_u$ states without breaking the interface structure\cite{Tang,PC}. Our energetic calculations show that the hydrogenation of C$_u$ yields an energy gain of 0.76 eV/H as compared to an energy cost of 0.95 eV/H for C$_b$. Therefore, we only consider the situation with the six C$_u$ fully hydrogenated as shown in Fig. 1(c) hereafter. Figure 1(d) plots the corresponding band structure without taking spin-polarization into account. It can be seen that the C$_u$ states are decoupled from Si$_u$ states and shift to lower energy. While around the Fermi level, there appears a nearly flat band dominantly contributed by Si$_u$ DB state. According to the Stoner criterion\cite{AStoner,BStoner}, spontaneous spin-polarization might occur given the large density of states at the Fermi level. Indeed, the spin-polarized state is energetically more favorable by 75 meV than the non-spin-polarized one. Figure 1(e) shows the spin-resolved band structure, in which the originally spin-degenerate Si$_u$ DB bands fully split into an occupied (majority) and an unoccupied (minority) band with an energy gap of $\sim$0.17 eV, hence contributing to 1 $\mu_B$ local magnetic moment.

It is worth to note that although the above results are obtained for 6$H$-SiC, the same trend holds for either 2$H$- or 4$H$-SiC according to our calculations, and is also independent of the SiC bilayer stacking sequence. An important difference is the magnitude of the bandwidth and spin-spliting of Si-DB band, which significantly affects the magnetic coupling between local moments. Below, we will discuss this in detail.


\emph{Long-range magnetic order.}--For practical applications, the existence of local moment is not enough and the long-range ferromagnetic order must be present. In fact, the emergence of charge carriers is unavoidable in our system due to the doping effect from SiC substrate\cite{Ristein,arXivSiCdoping}, which will play a significant role in mediating the magnetic interactions as intensively studied in dilute magnetic semiconductors\cite{sato} and $d^{0}$ ferromagnets\cite{dev2008defect,lijia}.  Moreover, it is possible to precisely control the carrier type and concentration by applying gate voltage. Hence we next investigate the possible long-range magnetic order in our systems while including the effect of charge doping.

To investigate the magnetic coupling of the local magnetic moments, we have explored two collinear magnetic structures. One is ferromagnetic (FM), i.e., all the local moments have a parallel alignment as shown in Fig. 2(a). The other is collinearly antiferromagnetic (ColAFM) with the row-wise antiparallel alignment as shown in Fig. 2(b). The magnetic interaction thus can be conveniently mimicked by mapping their respective total energies (FM and ColAFM) to the Heisenberg model. Note that the ColAFM configuration should not be realistic on a triangular lattice due to the frustration\cite{GeoFrus,frustration1}. However, our test calculation on a more realistic coplanar N\'{e}el (120$^{\circ}$) configuration (See Fig. 2(c)) gives a negligible energy lowering, less than 0.3 meV/Si-DB. Given the huge computational cost for noncollinear calculations, only the ColAFM situation is considered to simulate the AFM order.

\begin{figure}[tbh]
  \centering
  \includegraphics[width=0.75\textwidth]{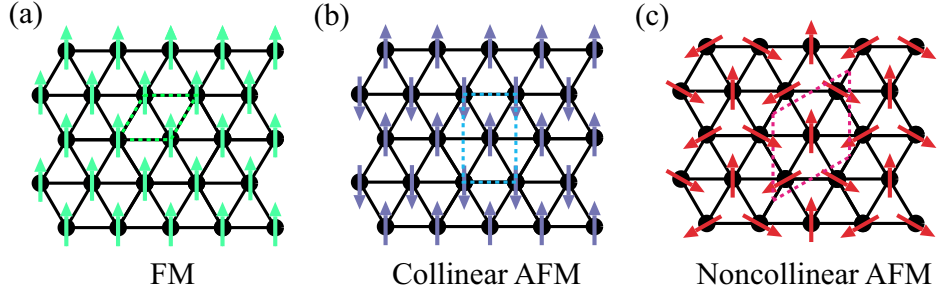}\\
  \caption{(Color online) Schematics of the calculated magnetic configurations: (a) Ferromagnetic (FM), (b) Collinear Antiferromagnetic (ColAFM) with the row-wise antiparallel structure, and (c) Coplanar noncollinear N\'{e}el (120$^{\circ}$) structure. Dash rhombuses are the unit cells.}\label{model}
\end{figure}

Figure 3(a) summarizes the energy difference ($\Delta E$) between the AFM and FM states as a function of carrier density ($\rho$) in hydrogenated epitaxial graphene on different hexagonal polytypes of SiC (2$H$, 4$H$ and 6$H$). For 4$H$-SiC, two kinds of stacking sequence are calculated as schematically shown in Fig. 3(b). Note that the charge doping range is comparable with the intrinsic doping level of epitaxial graphene\cite{Sun} and also experimentally accessible in hydrogenated epitaxial graphene \cite{GiesbersPRL}. The largest doping situation corresponds to a carrier density of 0.1 e/Si-DB ($\sim$4$\times$10$^{13}$/cm$^{2}$). Several features of ($\Delta E$) curves are observed in Fig. 3(a).

\begin{figure}[tbh]
  \centering
  \includegraphics[width=0.75\textwidth]{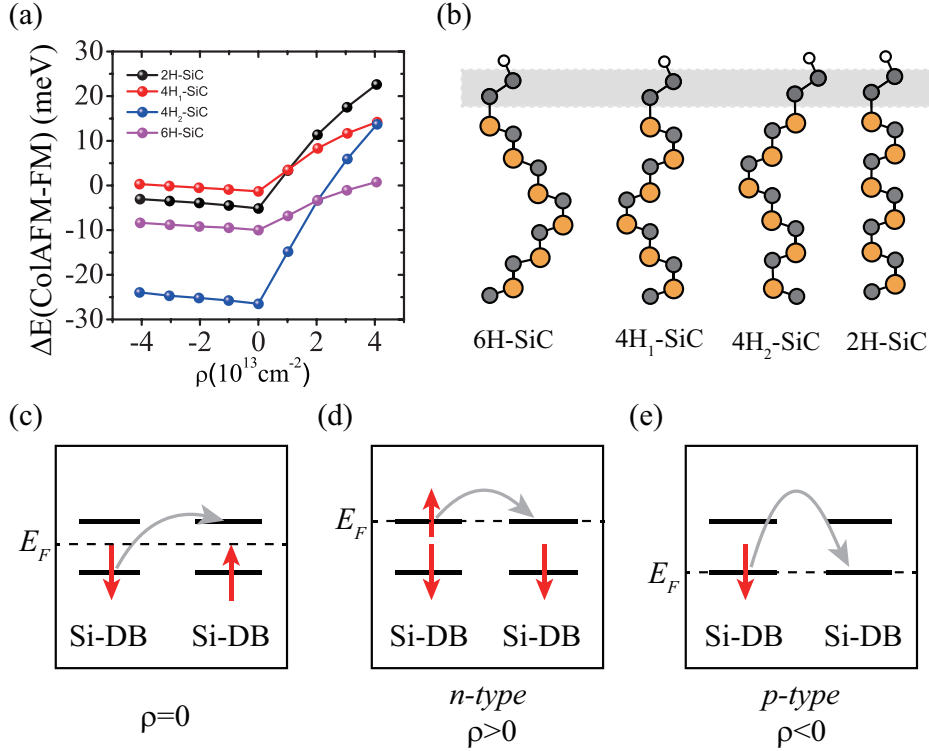}\\
  \caption{ (Color online) (a) Energy difference ($\Delta E$) between the AFM and FM states as a function of carrier density ($\rho$) in hydrogenated graphene on different SiC substrates. (b) Schematics of the SiC bilayer stacking sequence. The shaded area denotes graphene layer. (c)-(e) Schematics of the exchange mechanism without and with carrier doping. For the neutral system, virtual hopping yields an antiparallel alignment of nearest-neighbor local moments (c). In contrast, hopping of doped electrons (d) or holes (e) mediates a parallel alignment of local moments.}\label{model}
\end{figure}

Firstly, the AFM state has the lower energy under the ideal situation without any carrier doping independent of SiC polytypes. This can be understood in terms of Fig. 3(c). The Si-DB state is fully occupied in the majority spin and the virtual hopping is allowed when the nearest-neighboring local moments are in antiparallel alignment but forbidden for the parallel alignment configuration. Therefore, the ground state is AFM at $\rho =0$.

Secondly, as $|\rho|$ increases, the system becomes ``more'' ferromagnetic either for \emph{n}-type ($\rho$$>$0) or \emph{p}-type doping ($\rho$$<$0). The corresponding physics is illustrated in Figs. 3(d) and 3(e), revealing a typical double-exchange mechanism. Only when all the spins orient in parallel, the doped electrons/holes can move freely from one site to another, consequently reducing the system energy. Note that it is the competition between the double exchange and antiferromagnetic virtual hopping that determines the system ground state. Thus there exists a critical $\rho$, over which the $\Delta E$ becomes positive and the system undergoes a phase transition from the AFM to FM state. The asymmetric behavior between electron and hole indicates the different intra-site hopping interaction of varied carrier types.

Finally, there exists an interesting linear dependence of $\Delta E$ on $\rho$ regardless of SiC polytype, stacking sequence and carrier type. But different SiC substrates exhibit the different slops and intercept ($\rho=0$), showing their distinct effects on magnetic interaction. For example,  the parallel coupling is enhanced more rapidly in 2$H$- and 4$H_2$-SiC as the $\rho$ increases in comparison with 4$H_1$- and 6$H$-SiC. At $\rho=0$, the $\Delta E$ increases in the order of 4$H_2$-, 6$H$-, 2$H$ and 4$H_1$-SiC. These clearly show that SiC substrates, not only the polytypes but also the stacking sequence, can be used as an efficient way to modulate the long-range magnetic interaction in such a system.

In fact, the system can be described by the classical non-degenerate Hubbard model
\begin{equation}\label{1}
H=-\Sigma t_{\rm ij}c_{i\sigma}^{\dag}c_{j\sigma}+U\Sigma n_{i\uparrow}n_{i\downarrow} ,
\end{equation}
where the first term denotes the hopping ($t_{ij}$) interaction of local electrons from site $j$ to $i$, and the second term denotes the on-site Coulomb repulsion ($U$) with $n_{i\sigma}=c_{i\sigma}^{\dag}c_{i\sigma}$. Without carrier doping, the AFM exchange arises from virtual hopping of an electron with spin $\sigma$ to a neighboring site occupied with an electron of spin -$\sigma$(see Fig. 3(c)). The exchange energy is $-2t^{2}/U$ with the hopping matrix element $t_{ij}=t$. While in the system with a certain amount of carriers, their hopping from site to site with the matrix element $t_{\rm eff}$ would lead to the formation of the energy band:
\begin{equation}\label{1}
e(k)=-2t_{\rm eff}[ \cos(\frac{k_{x}}{2}+\frac{\sqrt{3}k_{y}}{2})+\cos(\frac{k_{x}}{2}-\frac{\sqrt{3}k_{y}}{2})+\cos(k_{x})]
\end{equation}
for the FM triangular lattice (Fig. 2(a)) and
  \begin{equation}\label{1}
e(k)=-2t_{\rm eff}\cos(k_{x})
\end{equation}
for the ColAFM configuration (Fig. 2(b)). Note that the carrier hopping is only allowed between the sites with parallel spins. When the $|\rho|$ is relatively small, just the lowest energy level would be occupied, thus approximately yielding the energy gains of $E_{\rm FM}=-6t_{\rm eff}$$|\rho|$ and $E_{\rm ColAF}=-2t_{\rm eff}$$|\rho|$, respectively. Further taking the virtual hopping contribution into account, we then have
\begin{equation}\label{1}
\Delta E=-2t^{\rm 2}/U+4t_{\rm eff} |\rho|.
\end{equation}

It clearly reveals the linear dependence of $\Delta E$ on $\rho$ in Eq. (4). At $\rho=0$, the $\Delta E$ corresponds to -2$t^2/U$. In fact, $t$ and $U$ can be estimated as the bandwidth and spin splitting of Si-DB band as shown in Figs. 1(d) and 1(e), respectively. The derived $t$ ($U$) values are 26.6, 37.0, 47.6, and 63.0 meV (688, 670, 520, and 562 meV) for 4$H_1$-, 2$H$-, 6$H$- and 4$H_2$-SiC. It yields the value of 2.76, 5.52, 10.1, and 17.2 meV for 2$t^2/U$, in good agreement with the data obtained from DFT calculations as shown in Fig. 3(a). Note that the ratio of $U/t$ is at least 9 to 1, and therefore, the pristine hydrogenated system should be characterized as a Mott insulator.

Doping charge carrier in Mott insulator will significantly alter its magnetic order and transport properties, leading to unconventional quantum phenomena such as magnetic phase transition\cite{Tokura} and high-temperature superconductivity\cite{RevModPhys.78.17}. The slope $t_{\rm eff}$ in Eq. (4) represents the energy gain due to hopping interaction of carriers. By fitting the data in Fig. 3(a) linearly, we obtain three typical values of $t_{\rm eff}$, namely, $\sim$2 meV, $\sim$25 meV and $\sim$60 meV. The value for hole doping is nearly universal ($\sim$2 meV) while there exist two typical values for electron doping on 4$H_1$- and 6$H$- ($\sim$25 meV) vs. 4$H_2$- and 2$H$-SiC ($\sim$60 meV). The distinct behaviors of electron and hole doping are generally found in doped Mott insulators, however, its physical origin is still under debate.\cite{RevModPhys.78.17}
In this case, the ferromagnetic exchange energy between two Si-DB spins can be evaluated by the form of double exchange interaction\cite{khomskii1997interplay}: t$_{hyb}^{4}$/$\Delta^{3}$, where t$_{hyb}$ is a hybridization matrix element between the local Si-DB states and surrounding charge densities which connect the inter-site Si-DB states, and $\Delta$ is the energy difference between Fermi level and energy level of surrounding charge densities. The electron-hole asymmetry might lie on the different Fermi level and the hybridization interactions. Anyway, the exact dependence of related parameters on the SiC stacking deserves to be further explored.


It is found that except the SiC polytype, the $\Delta E$ also depends on the SiC bilayer stacking sequence as demonstrated in the comparison between 4$H_1$ and 4$H_2$ cases (See Fig. 3(a)). The trend is also observed on 6$H$-SiC (not shown here). According to the previous studies\cite{Ristein,arXivSiCdoping}, the different stacking faults (i.e., an altered stacking sequence with respect to that in ideal cubic SiC) cause an entirely different local electrostatic environments at the SiC surface. This might be the underlying physics responsible for the modulation of interfacial magnetic interaction via SiC substrate.

\emph{Discussion.}--It should be emphasized that the recently reported room-temperature $d^{0}$ ferromagnetism in hydrogenated epitaxial graphene\cite{GiesbersPRL} is essentially different. Therein, an additional graphene overlayer, locating on the first graphene layer, is necessary for the long-range FM order, which, however, does not exist in our system. Nevertheless, our work can still provide some useful insights for the understanding of the experimental observations, e.g., the origin of local moments. More specially, Giesbers \emph{et al.}\cite{GiesbersPRL} unambiguously clarified the absence of FM signal in the samples with only the first graphene layer hydrogenated on 6$H$-SiC. Indeed, our results shown in Fig. 3(a) indicate that the AFM state is more favorable in energy than the FM state at the experimental doping level (0.9 $\mu_B$/hexagon projected area). This is well consistent with the experimental measurements\cite{GiesbersPRL}. Whereas, we predict that the FM order can be achieved by the change of SiC substrate. For example, the $\Delta E$ increases up to 22.6 meV at the same doping level on 2$H$-SiC. To estimate $T_c$, we employ the effective exchange Hamiltonian,  $H=-J_{\rm eff}\sum\limits_{\langle i, j\rangle}S_{i}S_{j}$, where $J_{\rm eff}$ is the effective nearest-neighbor exchange coupling and can be extracted from $\Delta E$, the $\langle i, j\rangle$ denotes nearest-neighbor sites and $S_{i}$ is the spin value at the site $i$. Under the mean field approximation, we can calculate the $T_c$ through the formula $T_c = \frac{2}{3k_B}zS(S+1)J_{\rm eff}$, where $z=6$ for the triangular lattice, $S=\frac{1}{2}$, and $k_B$ is the Boltzmann constant. It gives a $T_c$ of 394 K. Combining the high $T_c$ and inherent protection by atop graphene layer, the $d^{0}$ ferromagnetism on SiC should be a promising system for future spintronic applications.

In conclusion, we have proposed a new route to engineer $d^{0}$ ferromagnetism with high $T_c$ via the modulation of SiC bilayer stacking sequence using the first-principles calculations. It is found that hydrogenation always recovers the spin-splitting of Si-DB state while its local environment strongly depends on the SiC bilayer stacking sequence. By electron doping, the long-range ferromagnetism can be readily established. These new results call for experimental verification.

We acknowledge the support of the Ministry of Science and Technology of China  (Grant Nos. 2011CB921901 and 2011CB606405), and the National Natural Science Foundation of China (Grant Nos. 11304053 and 11334006).
\providecommand{\noopsort}[1]{}\providecommand{\singleletter}[1]{#1}%

\end{document}